\documentclass[amssymb,aps,twocolumn,showpacs,nofootinbib]{revtex4-1}
\usepackage[]{graphicx}
\usepackage[]{amsmath}
\usepackage{hyperref}
\usepackage{color}

\def\ket#1{| #1 \rangle}

\def\cA{\mathcal{A}}
\def\cB{\mathcal{B}}
\def\cC{\mathcal{C}}

\def\cF{\mathcal{G}}

\def\cL{\mathcal{L}}

\def\cS{\mathcal{S}}
\def\cT{\mathcal{T}}

\def\eq#1{Eq.~\eqref{eq:#1}}

\def\eq#1{Eq.~\eqref{eq:#1}}

\begin{document}

\title{Fault-tolerant conversion between the Steane and Reed-Muller quantum codes}
\author{Jonas T. Anderson}
\affiliation{D\'epartement de Physique, Universit\'e de Sherbrooke, Sherbrooke, Qu\'ebec, Canada}
\author{Guillaume Duclos-Cianci}
\affiliation{D\'epartement de Physique, Universit\'e de Sherbrooke, Sherbrooke, Qu\'ebec, Canada}
\author{David Poulin}
\email{David.Poulin@USherbrooke.ca}
\affiliation{D\'epartement de Physique, Universit\'e de Sherbrooke, Sherbrooke, Qu\'ebec, Canada}

\date{\today}

\begin{abstract}
Steane's 7-qubit quantum error-correcting code admits a set of fault-tolerant gates that generate the Clifford group, which in itself is not universal for quantum computation. The 15-qubit Reed-Muller code also does not admit a universal fault-tolerant gate set but possesses fault-tolerant $T$ and control-control-$Z$ gates. Combined with the Clifford group, either of these two gates generate a universal set. Here, we combine these two features by demonstrating how to fault-tolerantly convert between these two codes, providing a new method to realize universal fault-tolerant quantum computation. One interpretation of our result is that both codes correspond to the same subsystem code in different gauges. Our scheme extends to the entire family of quantum Reed-Muller codes.
\end{abstract}


\maketitle

One of the prominent techniques of fault-tolerant quantum computation is the use of transversal gates \cite{Sho96a}. In an architecture where each logical qubit is encoded in a code block which can protect against up to $t$ errors, a gate is said to be transversal if it does not couple qubits inside a given code block. As a consequence of transversality, the number of errors or faults in a block cannot increase under the application of a gate: the number of errors after the application of a gate is at most the number of initial errors on the data plus the number of faults in the execution of the gate. Single-qubit errors can propagate to single-qubit errors in other blocks, but these will be corrected independently on each block. In this way, an error-rate $\epsilon$ becomes $c\epsilon^{t+1}$ after error-correction, where $c$ is at most the number of different ways of getting $t+1$ faults in a single block. Recursing this procedure leads to the celebrated accuracy threshold theorem \cite{Sho96a,AB96a,Kit97b,KLZ98b,Pre98a}.

Unfortunately, it is not possible to construct a quantum code which admits a universal set of transversal gates \cite{EK09a}, so additional techniques are required. In many circumstances it is possible to fault-tolerantly implement the Clifford group, a finite sub-group of the unitary group which is not universal. 
In particular, all codes of the CSS family  have transversal controlled-not operations \cite{Ste99b}, and code deformation can be used to implement the entire Clifford group in topological codes \cite{BM09a}. 
Magic-state distillation and injection \cite{BK05b} is the most common technique  to complete the universal gate set. 

Recently, other techniques have been proposed to circumvent this no-go on transversal gates. Jochym-O'Conner and Laflamme \cite{JL13a} used a ``relaxed" notion of transversality which only demands that gates do not transform a single error or fault into an uncorrectable error, without prohibiting that it couples qubits from the same block. The same idea is responsible for the success of code deformation \cite{B10a1,BM09a}, which changes the error-correcting code in such a way that a full cycle returning to the original code  implements a gate. Because each step in the deformation acts on a number of qubits which is less than the minimum distance of  the codes, the transformation is fault-tolerant  despite being non-transversal \cite{BDP12a}. Schemes for topological quantum computation \cite{RH07a} are a form of code deformation. Paetznick and Reichardt \cite{PR13a} (see also a related idea of  Knill, Laflamme, and Zurek \cite{KLZ96a}) have proposed a scheme where transversal gates take the system outside the code space, but a subsequent round of error correction restores it. As we discuss below, this is conceptually equivalent to Bomb\'in's scheme  \cite{B13a} where transversal gates are applied to a subsystem codes \cite{KLP05a,Pou05b}, altering the gauge degree of freedom while applying a logical gate to the encoded data. The gauge is then returned to a standard state before a new gate is applied.

Here, we propose a scheme that converts between two codes which, jointly, possess a universal set of transversal gates. Clifford group transformations are realized in Steane's 7-qubit code \cite{Ste96a}, while the $T$ gate and/or the control-control-$Z$ gate are realized using the 15-qubit Reed-Muller code \cite{KLZ96a}; either of these last two gates is sufficient to complete the universal gate set, but an over-complete set can reduce the compilation overhead.  While it is always possible to convert between codes by preparing a special ancillary entangled state to teleport the data, our main contribution is a fault-tolerant scheme which directly converts the information in place. Much like in the approaches outlined above, the code is modified during the computation. One important difference here is that the codes involved have different numbers of qubits, an aspect that should be taken into account when optimizing resources to realize a given quantum circuit.  Similarly to the proposals of \cite{PR13a} and \cite{B13a}, our scheme can be seen as a subsystem encoding \cite{KLP05a,Pou05b} with different gauge fixing. In fact, our approach should be seen as a generalization of \cite{PR13a}, which enables a much richer set of fault-tolerant gates and extends to the entire quantum Reed-Muller code family.  

The rest of this paper is organized as follows. After a brief review of classical and quantum codes, we present the family of quantum Reed-Muller codes and highlight some of their key properties. Then, we review transversal gate constructions for these codes, focusing in particular on the first two instances of the family which correspond to Steane's 7-qubit code and a 15-qubit Reed-Muller code. We then explain the conversion scheme, which essentially relies on a recursive definition of the Reed-Muller codes. Lastly, we present an alternative derivation in terms of subsystem codes, and conclude by discussing possible applications of our scheme.

\medskip
\noindent{\em Codes---}
An $n$-bit classical linear code encoding $k$ bits is defined as the null-space of a $(n-k)\times n$ parity-check matrix $H$ (in $\mathbb Z_2$ arithmetic), i.e. $\cC = \{x \in \mathbb Z_2^n : H x = 0\}$.
Its minimum distance  $d$ is the minimum number of bit-flips required to map one code-word to another. Given an erroneous string $x' = x+e$ obtained from a code word $x$ and error $e$, the error syndrome is given by $s = Hx' = He$ and can unambiguously identify any error acting on less than $(d-1)/2$ bits. The code can also be defined as the row-space of a $k\times n$ generator matrix $G$, i.e. $\cC = {\rm row}(G)$, which is the dual of $H$, meaning that it is a matrix of largest rank which obeys $HG^T = 0$. 

A stabilizer code encoding $k$ qubits into $n$ qubits is specified by a set $\cA$ of $n-k$ independent stabilizer generators, which are commuting and hermitian elements of the $n$-qubit Pauli group (obtained from $n$-fold tensor product of the $2\times 2$ identity $I$ and the Pauli matrices $X$, $Y$, and $Z$). The code space $\cC$ is a subspace of the $n$-qubit Hilbert space stabilized by $\cA$:  
\begin{equation}
\cC = \{\ket\psi \in (\mathbb C^2)^{\otimes n} : A \ket\psi = \ket\psi \quad \forall A\in \cA\}.
\end{equation}
Equivalently, it can be defined as the image of the code projector $P_\cA = \prod_{A\in \cA} \frac{I+A}2 = \frac1{2^{|\cA|}}\sum_{S\in \cS} S$ where $\cS$ is the stabilizer group generated by $\cA$. 
When a code state $\ket\psi\in \cC$ undergoes a Pauli error $E$, error correction is realized by measuring the stabilizer generators. The $\pm 1$ measurement outcome of measuring $A_j\in \cA$ indicates whether $A_j$ commutes or anti-commutes with $E$: $A_j (E\ket\psi = \pm EA_j\ket\psi = \pm (A\ket\psi)$.  Logical operators transform the state but preserve the code space, i.e. they are  elements of $N(\cS)-\cS$, where $N$ denotes the normalizer of a group. A code has distance $d$ if it takes an error of weight $d$ or more to map a codeword to a distinct codeword. These parameters of a  code are collectively denoted $(n,k,d)$ in the classical setting and $[[n,k,d]]$ in the quantum setting. 

\medskip
\noindent{\em The Reed-Muller code---}
The Reed-Muller codes of order 1 can be defined recursively  \cite{MS77a}: the code ${\sf RM}(1,1)$ has generator matrix
\begin{equation}
G_1 = \left(\begin{array}{cc} 1& 1 \\ 0&1 \end{array}\right)
\end{equation}
and the code ${\sf RM}(1,m)$ has generator matrix
\begin{equation}
G_{m+1} = \left(\begin{array}{cc} G_m & G_m \\ 0\ldots 0&1\ldots 1 \end{array}\right).
\end{equation}
The dual of ${\sf RM}(1,m)$ is ${\sf RM}(m-2,m)$ and has generator matrix $H_m$. Quantum codes are derived from shortened Reed-Muller codes $\overline{{\sf RM}}(1,m)$, where the first row and column are deleted from $G_m$. We can similarly define shortened dual codes $\overline{{\sf RM}}(m-2,m)$ with generator matrix $\overline{H}_m$. Hence, the generator matrices of $\overline{{\sf RM}}(1,m)$ obey the recursive definition
\begin{equation}
\overline G_{m+1} = \left(\begin{array}{ccc} 
 \overline G_m & \overline G_m & 0\\  0\ldots 0& 1\ldots 1 & 1  \end{array}\right)
 \label{eq:recursionRM}
\end{equation}
(we have permuted the columns for later convenience). Note that $\overline{{\sf RM}}(m-2,m)$ is not the dual of $\overline{{\sf RM}}(1,m)$. Using this definition, the following Facts can easily be verified (see Appendix A) by induction for $m\geq 2$:
\begin{enumerate}
\item For $x\in  {\sf RM}(1,m)$ or $\overline{{\sf RM}}(1,m)$, $|x| = 0\mod 2^{m-1}$. 
\item For $m\geq 3$, $\overline{{\sf RM}}(1,m)$ is contained in its dual, i.e., $\overline G_m\overline G_m^T = 0$.
\item The minimum distance of the dual code is $3$.
\item $\overline{{\sf RM}}(1,m)$ is contained in the dual of $\overline{{\sf RM}}(m-2,m)$, i.e. $\overline H_m \overline G_m^T = 0$.
\item $\overline{{\sf RM}}(1,m)$ is contained in $\overline{{\sf RM}}(m-2,m)$, i.e. ${\rm row}(\overline G_m) \subset {\rm row} (\overline H_m)$ .
\item For $x_1,x_2,\ldots x_k \in \overline{\sf RM}(1,m)$, $x_1\cdot x_2\cdot\ldots x_k = 0\mod 2^{m-k}$.
\end{enumerate}

The quantum Reed-Muller codes \cite{Ste99d} ${\sf QRM}(m)$ derived from $\overline{{\sf RM}}(1,m)$ codes are CSS codes, meaning that their stabilizer generators break into two sets $\cA^x_m$ and $\cA^z_m$ \cite{CS96a,Ste96b}. Elements of $\cA_m^x$ are obtained from rows of $\overline G_m$, by substituting 1s by $X$s and 0s for $I$s. Elements of $\cA_m^z$ are obtained in a similar way, but from the generator matrix of the shortened dual code $\overline{{\sf RM}}(m-2,m)$. But since $\overline{{\sf RM}}(1,m)\subset \overline{{\sf RM}}(m-2,m)$ (Fact 5), it follows that  $\cA^z_{m}$ contains the same operators as $\cA_{m}^x$ with $X$s replaced by $Z$s, plus some additional operators coming from the dual. In other words, if we define $ \cA_m^{\prime z} \equiv H^{\otimes n} \cA_m^x H^{\otimes n} $ as the $z$-stabilizers corresponding to the rows of $\overline G_m$, then $\cA_m^z = \cA_m^{\prime z} \cup \tilde\cA_m^{z}$ for some set of $z$-stabilizer generators $\tilde\cA_m^z$.  

In a CSS code, $\cA^x$ detects $z$-type errors and $\cA^z$ detects $x$-type errors. Since we have defined the stabilizers of the quantum code in terms of the generator matrix of the classical code, it follows that the minimum distance of the quantum code is given by the minimum distance of the dual classical code, which is $d=3$ (Fact 3), so it can correct any single qubit error. Since the shortening procedure has removed one stabilizer from the original code and one from the dual code, it follows that the parameters of the code are $[[n=2^m-1, k=1, d=3]]$. The logical operators are given by the rows that were removed in the shortening procedure, they are $\overline X_m = X^{\otimes n}$ and $\overline Z_m = Z^{\otimes n}$. Finally, note that the commutation of the stabilizer generators follows from the orthogonality of $\overline{{\sf RM}}(1,m)$ and $\overline{{\sf RM}}(m-2,m)$ (Fact 4). 

\medskip
\noindent{\em Transversal gates ---}
The logical $0$ state of a code should be a simultaneous +1 eigenstate of $\overline Z$ and all elements of $\cA$. The state $\ket{{\bf 0}}$ (we use the bold symbols {\bf 0} and {\bf 1} to designate respectively strings of 0s and 1s of lengths fixed by the context) is already a +1 eigenstate of $\overline Z$ and of all $\cA_m^z$, so we obtain the logical 0 by projecting it onto the +1 eigenspace of elements of $\cA_m^x$:
\begin{align}
\ket{\overline 0}_S &= \prod_{A\in \cA_m^x} \frac{I+A}2 \ket{\bf 0}\\
&= \frac 1{2^n} \sum_{S \in \cS_m^x} S \ket{\bf 0}\\
&= \frac 1{2^n} \sum_{x \in {\rm row}(\overline G_m)} \ket{x}.
\end{align}
The logical 1 is obtained by applying $\overline X_m$ to this state, so it is $\ket{\overline 1}= \frac 1{2^n} \sum_{x \in {\rm row}(\overline G_m)} \ket{ x \oplus \bf{1}}$. It follows from Fact 1 that $\ket{\overline 0}$ is the superposition of strings of weight $0\mod 2^{m-1}$ and $\ket{\overline 1}$ is the superposition of strings of weight $-1\mod 2^{m-1}$.

Consider now the single-qubit gate $Z(\omega_\ell) = {\rm diag}(1,\omega_\ell)$ where $\omega_\ell$ is the $\ell$th root of unity. Observe that for any $n$-bit string $x$, $Z(\omega_\ell)^{\otimes n}\ket x = \omega_\ell ^{|x|} \ket x = \omega_\ell ^{(|x|\mod \ell)} \ket x$. From the above consideration on the weights of the basis states appearing in the logical states $\ket{\overline 0}$ and $\ket{\overline 1}$,  it follows that for $\ell = 2^{m-1}$, the transversal gate $Z(\omega_\ell)^{\otimes n}$ acts as the logical $Z(\omega_\ell)^\dagger$ on ${\sf QRM}(m)$, so it is transversal \cite{CAB12a,BH12b,LC13a}.

The codes ${\sf QRM}(m)$ also have a transversal $k$-fold controlled-$Z$ gate for $k\leq m-2$. Note that the transversal $k$-fold controlled gate acts on a basis state $\ket{x_1}\ket{x_2}\ldots\ket{x_{k+1}}$ by introduction of a phase factor  $(-1)^{x_1\cdot x_2\cdot\ldots x_{k+1}}$. A logical state $\ket{\bar y}$ is the superposition of states of the form $\ket{x+y\bf 1}$ where $x\in \overline{\sf RM}(1,m)$. When acted on by a transversal $k$-fold controlled-$Z$ gate, a logical state $\ket{\bar y_1}\ket{\bar y_2}\ldots \ket{\bar y_{k+1}}$ will pick up a phase factor  $(x_1+y_1{\bf 1})\cdot (x_2+y_2{\bf 1})\cdot\ldots (x_{k+1}+y_{k+1}{\bf 1})$ where  $x_j\in \overline{\sf RM}(1,m)$ for all j. Expanding this product, all terms containing $x$s produce a trivial phase due to Fact 6, so only the term $y_1y_2\ldots y_{k+1}$ contributes to the phase which produces the desired transformation.

The 7-qubit Steane code is derived from the classical code $\overline{{\sf RM}}(1,3)$, a.k.a. the classical (7,4,3) Hamming code. This is a special case as it is self-dual, which implies that $\cA_3^x$ and $\cA_3^z$ are equal up to exchanging $X$s for $Z$s. As a consequence it has transversal Clifford gates.  The Hadamard gate $H$ exchanges $X$ and $Z$. It is thus clear that the transversal gate $H^{\otimes 7}$ preserves the code space (as it only swaps $\cA_3^x$ with $\cA_3^z$) and acts as the logical Hadamard by exchanging $\overline X$ with $\overline Z$. The CNOT acting on two qubits maps the operators ($IX$, $XI$, $IZ$, $ZI$)  to ($IX$, $XX$, $ZZ$, $ZI$). The transversal gate CNOT$^{\otimes 7}$ therefore acts on the logical operators as a logical CNOT, and maps the set of generators $\{ I\cA_3^x,\cA_3^xI,I\cA_3^z,\cA_3^zI\}$ of $\cS_3\otimes \cS_3$ to $\{ I\cA_3^x, \cA_3^x\cA_3^x, \cA_3^z\cA_3^z, \cA_3^zI\}$, which is simply a different set of generators for $\cS_3\otimes \cS_3$, so the code is preserved. Finally, the phase gate $P$ corresponds to $Z(\omega_4)$ defined above and is transversal as we have seen.

%The next code in the family is called the 15-qubit quantum Reed-Muller code, obtained from $\overline{{\sf RM}}(1,4)$. It has a transversal $T = Z(\omega_8)$ gate following the general argument above. It also has a transversal control-control-$Z$ gate \cite{PR13a}. To see this, note that the action of a transversal control-control-$Z$ gate on a computational basis state $\ket x\otimes\ket y\otimes \ket z$ returns the same state with the phase factor $(-1)^{|x\cdot y\cdot z|}$. Since $|x\cdot y| = 0\  {\rm mod}\  2$ and $|x\cdot y\cdot z| = 0 \ {\rm mod}\  2$ for all $x,y,z \in {\rm row}(\overline G_4)$, it follows that the phase factor is trivial for every code state except when $x, y$, and $z$ are in row$(\overline G_4)\oplus {\bf 1}$, corresponding to the action of the logical control-control-$Z$ gate.

\medskip
\noindent{\em Conversion---}
As we have seen above, the $z$-type stabilizers of quantum Reed-Muller codes can be broken into two sets $\cA_m^z = \cA_m^{\prime z} \cup \tilde\cA_m^{z}$ where $\cA_m^{\prime z}$, like $\cA_m^x$, is obtained from the rows of $\overline G_m$. It follows from \eq{recursionRM} that these can be defined recursively. Given two ordered sets $\cA = \{A_1,A_2,\ldots\}$ and $\cB = \{B_1,B_2,\ldots\}$, we introduce the notation $\cA\times \cB = \{A_1\otimes B_1, A_2\otimes B_2,\ldots\}$. Given this definition, it follows from \eq{recursionRM} that
\begin{align}
\cA_{m+1}^x &= \left\{
\begin{array}{lllll}
\cA_m^x& \times &\cA_m^x &\otimes &I,\\ 
I^{\otimes n} & \otimes & \overline X_m& \otimes &X
\end{array}\right\},\quad {\rm and} 
\label{eq:recursiveQRM1}\\
\cA_{m+1}^{\prime z} &= \left\{
\begin{array}{lllll}
\cA_m^{\prime z}& \times &\cA_m^{\prime z} &\otimes &I,\\ 
I^{\otimes n} & \otimes & \overline Z_m& \otimes &Z
\end{array}\right\}.
\label{eq:recursiveQRM2}
\end{align}
Our central result can be summarized by the observation that the stabilizers $\tilde \cA_m^z$ are not needed in order to correct  single-qubit errors. Since elements of $\cA_m^x$ can unambiguously discriminate all single-qubit $z$-errors, it follows that $\cA_m^{\prime z} = \cA_m^z - \tilde\cA_m^z$ can unambiguously discriminate all single-qubit $x$-errors. Operators from $\tilde \cA_m^z$ are superfluous.  Thus, starting from the ``relevant'' stabilizers $\cA_m^x$ and $\cA_m^{\prime z}$, there are many ways to complete the list of stabilizers in order to obtain a good error-correction code. Our scheme will make use of this freedom to convert between different codes. 
%In what follows, for concreteness, we focus on the relation between the first two codes in the family, $QRM(3)$ and $QRM(4)$.

Let us first explain how to convert from ${\sf QRM}(m)$ to ${\sf QRM}(m+1)$. We begin with some information encoded in an $(2^m-1)$-qubit state of  ${\sf QRM}(m)$, $\ket{\overline \psi}_m$. We prepare an $2^m$-qubit quantum state  $\ket{\Phi} = \frac 1{\sqrt 2}(\ket{\overline 0}_m\ket 0 + \ket{\overline 1}_m\ket 1)$  consisting of a maximally entangled state between a bare qubit and a qubit encoded in ${\sf RM}(m)$. Viewing the joint state $\ket{\overline\psi}_m\otimes \ket\Phi$ as an encoded state of a $(2^{m+1}-1)$- qubit code, we can write the generators for this ``extended quantum Reed-Muller code'' as
\begin{equation}
\begin{array}{cccccc}
\cA_m^z& \otimes & I^{\otimes n} &\otimes &I\\
\cA_m^x& \otimes & I^{\otimes n} &\otimes &I\\
I^{\otimes n}& \otimes &\cA_m^z &\otimes &I\\
I^{\otimes n}& \otimes &\cA_m^x &\otimes &I\\
I^{\otimes n} & \otimes & \overline Z_m& \otimes &Z\\
I^{\otimes n} & \otimes & \overline X_m& \otimes &X
\end{array}
\end{equation}
We can change the generating set without changing the code and instead use
\begin{equation}
\begin{array}{cccccc}
\cA_m^{\prime z}& \times  &\cA_m^{\prime z}&  \otimes &I\\
\cA_m^{x}& \times  &\cA_m^x&  \otimes &I\\
I^{\otimes n} & \otimes & \overline Z_m& \otimes &Z\\
I^{\otimes n} & \otimes & \overline X_m& \otimes &X\\
\tilde \cA_m^{z}& \times  &\tilde\cA_m^{z}&  \otimes &I\\
\cA_m^z&\otimes& I^{\otimes n} &\otimes &I\\
\cA_m^x&\otimes& I^{\otimes n} &\otimes &I
\end{array}
\label{eq:ExtSteane}
\end{equation}
We immediately recognize the first $2m+2$ generators of this list [first four rows of \eq{ExtSteane}] as generating the relevant stabilizers of ${\sf QRM}(m+1)$, i.e. $\cA_{m+1}^x$ and $\cA_{m+1}^{\prime z}$. Indeed, compare to Eqs.~(\ref{eq:recursiveQRM1},\ref{eq:recursiveQRM2}). Thus, only operators from the last three lines of  \eq{ExtSteane} differ, and must be substituted by $\tilde\cA_m^{z}$ to convert into ${\sf QRM}(m+1)$. In fact, only the $m$ stabilizers of the last line are a problem, since $\tilde \cA_m^{z} \times  \tilde\cA_m^{z}  \otimes I$ and   $\cA_m^z\otimes I^{\otimes n} \otimes I\subset \tilde \cA_{m+1}^z$.

But as explained in the previous paragraphs, these $m$ stabilizers are superfluous in the sense that they are not required to diagnose single-qubit errors. Thus, if we fault-tolerantly measure all stabilizers of ${\sf QRM}(m+1)$ on the state $\ket{\overline\psi}_m\otimes \ket\Phi$, we can use the syndrome from the first six rows of \eq{ExtSteane} to diagnose errors, and remove any syndrome associated to the last $m$ stabilizers by a fault-tolerant error-correction procedure (or by adapting the Pauli frame). Specifically, given a set of stabilizer generators $\cA = \{A_1, \ldots A_{n-k}\}$ and logical operators $\cL = \{\overline X_a,\ldots \overline X_k, \overline Z_1,\ldots \overline X_k\}$, there exists a set of ``pure errors'' $\cT = \{T_1,\ldots T_{n-k}\}$ such that $T_j$ commutes with all elements of $\cL$, $\cT$, and $\cA$ except $A_j$ with which it anti-commutes. The error-correction procedure alluded to above then simply consist in applying the operator $T_j$ when one of the last $m$ stabilizer $A_j$ reveals a syndrome $-1$. 

To summarize, to convert from ${\sf QRM}(m)$ to ${\sf QRM}(m+1)$, we first fault-tolerantly prepare the $2^m$-qubit stabilizer state $\ket\Phi$,  append it to the system, fault-tolerantly measure the stabilizer generators of ${\sf QRM}(m+1)$, error-correct given the first $2^{m+1}-m-2$ syndrome bits (first six rows of \eq{ExtSteane}) and restore the last $m$ syndrome bits using their associated pure errors. 

To convert from the ${\sf QRM}(m+1)$ to ${\sf QRM}(m)$, we simply fault-tolerantly measure the stabilizers of \eq{ExtSteane}, use the first $2^{m+1}-m-2$ syndrome bits (first six rows of \eq{ExtSteane}) to diagnose errors, and restore the last $m$ syndrome bits using the associate pure errors.  We can then remove the additional $2^m$ qubits and be left with the $(2^m-1)$-qubit state $\ket{\overline \psi}_m$ encoded in ${\sf QRM}(m)$.  

\medskip
\noindent{\em Subsystem code interpretation---}
It is possible to recast the above conversion scheme using the subsystem code formalism \cite{KLP05a,Pou05b}, which highlights its similarity with Paetznick and Reichardt \cite{PR13a} and Bomb\'in \cite{B13a} schemes. We can define a stabilizer code from the stabilizers that are common to ${\sf QRM}(m+1)$ and the extended ${\sf QRM}(m)$. There are $2^{m+1} - m-2$ of these and they are given by the first six lines of \eq{ExtSteane}. Thus, this code encodes $k=m+1$ logical qubits and has minimum distance $d=3$, so it can error-correct any single-qubit error. 

One of these logical qubits, which we label 0,  is the one encoded in the original code and has logical operators $\overline X^0 = \overline X_m$ and $\overline Z^0 = \overline Z_m$. The other logical operators associated to ``gauge qubits'', $\overline X^j$ with $j=1,\ldots, m$ correspond to elements of the last line of \eq{ExtSteane}. Their conjugate partners $\overline Z^j$ are generated by elements of $\tilde \cA_{m+1}^z$. 

We obtain a subsystem code by choosing to encode information only in the first logical qubit of the code. The other logical qubits $j=1,2,\ldots m$ carry no information, and can be fixed to an arbitrary state. The conversion scheme described above then simply consists in fixing these $m$ gauge qubits all in state $\ket{\overline 0}$ or all in state $\frac 1{\sqrt 2}(\ket{\overline 0}+\ket{\overline 1})$. The first scenario can be realized by measuring the operators $\overline Z^j$, and flipping the qubit using $\overline X^j$ if the outcome is $-1$. This procedure brings the state to the extended quantum Reed-Muller code, and the last $2^m$ qubits can be discarded to obtain a state encoded in ${\sf QRM}(m)$. The second scenario can be realized by measuring the operators $\overline X^j$, and flipping the qubit using $\overline Z^j$ if the outcome is $-1$. This procedure brings the state to ${\sf QRM}(m+1)$. 

Thus, we see that the different quantum Reed-Muller codes all correspond to the same subsystem code with different gauge fixing. Depending on the chosen gauge, some qubits become unentangled with the part of the code supporting the data, and can be discarded. At the bottom of this hierarchy is Steane's 7-qubit code, which realizes the entire Clifford group transversally. Above is an infinite family of quantum Reed-Muller codes which admit increasingly complex transversal gates.

\medskip
\noindent{\em Conclusion \& Outlook---}
We have presented a scheme to directly and fault-tolerantly convert between a family of quantum error correcting codes. By combining the transversal gate sets of these codes, we obtain an (over-complete) universal gate set. Our result offers a deeper understanding of a recent proposal \cite{PR13a} and extends it in many ways.

An important advantage of our conversion scheme is its potential reduction of overhead. The scheme of \cite{PR13a} requires $N_L15^\ell$ qubits to encode the logical state, where $N_L$ is the number of logical qubits and $\ell$ is the number of concatenations. Using our approach, this number becomes $(N_L-N_{NC})7^\ell + N_{NC}15^\ell$, where $N_{NC}$ is the maximum number of non-Clifford operations being executed at any given time in the algorithm. Unless these non-Clifford operations can be highly parallelized, the savings are considerable. We can envision an architecture where  special areas in the computer are dedicated to the execution of non-Clifford gates. In those areas, the encoding uses the Reed-Muller code, while the rest of the computer is encoded with Steane's code. Qubits are converted in and out of these areas to realize non-Clifford gates.

The Reed-Muller code family  can be used to distill magic states \cite{CAB12a,BH12b,LC13a,PR13a}. Distillation is a procedure which uses Clifford operations to increase the fidelity of non-stabilizer states, which can be injected in the computation to realize non-Clifford transformations \cite{BK05b}. Our scheme could potentially improve distillation procedures based on Reed-Muller codes since all Clifford operations could be performed on smaller codes. We leave the detailed study of this proposal for future work.

Finally, we note that the higher-order Reed-Muller codes ${\sf RM}(r,m)$  obey a similar recursive definition 
\begin{equation}
G_{r,m+1} = \left(\begin{array}{cc}
G_{r,m} & G_{r,m}\\
0 & G_{r-1,m}
\end{array}
\right)
\label{eq:RMrm}
\end{equation}
and are dual-containing when their rates is more than 1/2 \cite{MS77a}, so our conversion procedure  can be extended to this broader class of codes (see appendix B).

\medskip
\noindent{\em Acknowledgements---} This work was partially funded by Canada's NSERC, Qu\'ebec's FRQNT through the network INTRIQ, and the Lockheed Martin Corporation. DP acknowledges the hospitality of The University of Sydney where this project was completed.

\bibliographystyle{/Users/dpoulin/archive/hsiam}
\bibliography{/Users/dpoulin/archive/qubib}

\bigskip 
\noindent{\bf Appendix A---} 
In this appendix we prove the first 5 properties of the shortened Reed-Muller codes listed in the main text as Facts. Fact 6 will be proved in Appendix B. It will be useful to make use of an alternative recursive definition of these codes \cite{MS77a}:
\begin{equation}
{\sf RM}(1,m+1) = \left\{(x,x), (x,x+{\bf 1}) : x \in {\sf RM}(1,m)\right\}.
\label{eq:recursiveCode}
\end{equation}

Fact 1. For ${\sf RM}(1,m)$, the base case $m=2$ can be verified directly. Suppose that the fact holds for $m$, which means that the allowed weights of elements of ${\sf RM}(1,m)$ are $w_m = 0$, $2^{m-1}$, or $2^m$. Using \eq{recursiveCode}, we see that the the weight of elements of ${\sf RM}(1,m+1)$ will be either $2w_m$ or $w_m + (2^m-w_m)$, so the condition is satisfied. When we shorten the code to get $\overline{{\sf RM}}(1,m)$, we remove a row from $G_m$ which contains all 1s and then remove a column containing all 0s. Thus, we have 
\begin{equation}
{\sf RM}(1,m) = \left\{(0,x), (1,x+{\bf 1}) : x \in \overline{{\sf RM}}(1,m)\right\}.
\label{eq:shortening}
\end{equation}
Thus, the set $\{ (0,x) : x\in \overline{{\sf RM}}(1,m)\}$ is a subset of ${\sf RM}(1,m)$, so the property holds for $\overline{{\sf RM}}(1,m)$ as well.

Fact 2. The base case $m=3$ is well known, it corresponds to the Hamming code (Steane's code). The induction yields
\begin{equation}
\overline G_{m+1}\overline G_{m+1}^T = 
\left(
\begin{array}{cc}
0 & \overline G_m \cdot {\bf 1}^T \\
{\bf 1} \cdot \overline G_m^T & 0
\end{array}\right). 
\end{equation}
Noting that $\overline G_m \cdot {\bf 1}^T$ is simply the vector of weights $\mod 2$ of the rows of $\overline G_m$ and that these are even by Fact 1 proves Fact 2.

Fact 3. Since we are interested in the dual code, we should think of $G_m$ as the parity check matrix of a code. The base case $m=2$ corresponds to the parity-check matrix
\begin{equation}
\overline G_2 = \left(\begin{array}{ccc}
1&0&1\\
0&1&1
\end{array}\right)\label{eq:G2}.
\end{equation}
The minimum distance is obviously bounded by the length of the code $d\leq 3$. This parity-check matrix can uniquely identify any single bit error since all its columns are distinct, so it has minimum-distance 3. In its recursive definition \eq{recursionRM}, $\overline G_{m+1}$ contains three blocks of bits: the first two of size $2^m-1$ and the last of size 1. It is clear that the minimum distance for $m+1$ is no greater than the minimum distance for $m$, since an error occurring in the first block is only seen by $\overline G_m$. On the other hand, a single-bit error occurring in different blocks will trigger different syndrome patterns. If it is block 1 its first $m$ syndrome bits will be non-trivial and its last syndrome bit will be trivial. If it is block 2 its first $m$ syndrome bits will be non-trivial and its last syndrome bit will be non-trivial. If it is block 3 its first $m$ syndrome bits will be trivial and its last syndrome bit will be non-trivial. Moreover, in each case the syndrome can uniquely identify the error by induction, proving the fact.

Fact 4. To prove this fact it is important to know that for shortening,  the row which is deleted from $H_m$ is all 1s and that the subsequently deleted column is all 0s. The fact that $H_m$ contains an all 1s row simply reflects the fact that elements of ${\sf RM}(1,m)$ have even weight for $m\geq 2$. The fact that the rest of the first column is all 0s can always be obtained by Gaussian elimination.  By definition, $G_mH_m^T = 0$, or in other words
\begin{align}
\left(
\begin{array}{cc}
1&1\ldots 1\\
0& \\
0 & \overline G_m \\
0 & 
\end{array}\right)
\left(
\begin{array}{cc}
1&0\ldots 0\\
1& \\
1& \overline H_m^T \\
1 & 
\end{array}\right) \\ =
\left(
\begin{array}{cc}
0&\overline H_m\cdot {\bf 1}^T\\
{\bf 1}\cdot \overline H_m^T& \overline G_m\overline H_m^T 
\end{array}\right) 
= 0.
\end{align}

Fact 5. First, we prove that ${\rm row}(G_m) \subset {\rm row}(H_m)$. This follows from the fact that $G_mG_m^T = 0$, which we prove by induction:
\begin{equation}
 G_{m+1} G_{m+1}^T = 
\left(
\begin{array}{cc}
0 &  G_m \cdot {\bf 1}^T \\
{\bf 1} \cdot  G_m^T & 0
\end{array}\right). 
\end{equation}
The r.h.s is 0 since rows of $G_m$ have even weight from Fact 1. The fact follows from the observation that $\overline G_m$ and $\overline H_m$ are obtained from $G_m$ and $H_m$ by the same shortening procedure: first remove an all 1s row and then remove an all 0s column. 

Fact 6: For this Fact it is convenient to define ${\sf RM}(1,m)$ as boolean polynomials with all terms of degree 1 \cite{MS77a}. Then, $x_1\cdot x_2\cdot \ldots x_k$ is a boolean polynomial with all terms of degree $k$, and these have  weights $0 \mod 2^{m-k}$ \cite{MS77a}.

\medskip
\noindent{\bf Appendix B---} 
In this appendix we discuss the generalization to higher rank Reed-Muller codes, defined recursively by \cite{MS77a}
\begin{align}
&{\sf RM}(r,m+1) = \\ &\left\{(x,x+y) : x \in {\sf RM}(r,m), y \in {\sf RM}(r-1,m)\right\}, \nonumber
\label{eq:recursiveCode_rm}
\end{align}
or equivalently by \eq{RMrm}.

Denote $G_{r,m}$ the generator matrix of ${\sf RM}(r,m)$. We choose a pair of codes  ${\sf RM}(m-r-1,m)$ and ${\sf RM}(m-r,m+1)$ both of rates greater than $1/2$. Such codes contain their dual  \cite{MS77a}, so in particular the first code contains ${\sf RM}(r,m)$, which implies that ${\sf RM}(r,m)$ is self-orthogonal, and the same reasoning applies to ${\sf RM}(r,m+1)$. Since ${\sf RM}(m-r,m)$ has a rate greater than ${\sf RM}(m-r-1,m)$, it follows that ${\sf RM}(r-1,m)$ is also self-orthogonal. In short, we have just shown $G_{r,m}G_{r,m}^T = 0$, $G_{r-1,m}G_{r-1,m}^T = 0$, and $G_{r,m+1}G_{r,m+1}^T = 0$. This last equality combined to \eq{RMrm} implies that $G_{r,m}G_{r-1,m}^T = 0$. 

As a consequence of these orthogonality conditions, we can use the rows of $G_{r,m}$ to build a self-dual CSS code ${\sf QRM}(r,m)$. Similarly, we can build a self-dual CSS code $\cF$ from the union of the rows of ${\sf RM}(r,m)$ and ${\sf RM}(r-1,m)$. The code $\cF$ has minimum distance $\geq d_{r-1,m}$. There are many inequivalent ways of building subsystem codes from these, by converting some logical qubits into gauge qubits, by shortening the codes, and by adding additional stabilizers $\tilde\cA_{r,m}$ or equivalently fixing the gauge in various ways. Below we briefly discuss one possible construction, which converts between two subsystem codes with stabilizers given by ${\sf QRM}(r,m)$ and ${\sf QRM}(r,m+1)$, and can tolerate $d_{r,m} = 2^{m-r}$ single-qubit errors. 

$m+1 \rightarrow m$ conversion: We begin in a subsystem code with stabilizers ${\sf QRM}(r,m+1)$. As of \eq{RMrm}, we can naturally partition the $2^{m+1}$ qubits into two blocks of $2^m$ qubits. We can measure the stabilizers of $\cF$ on the second block, and correct any errors it reveals. This leaves the first block in the code ${\sf QRM}(r,m)$. The logical operators of ${\sf QRM}(r,m)$ acting on the first block are preserved by this procedure.

$m \rightarrow m+1$ conversion: We begin in the stabilizer code ${\sf QRM}(r,m+1)$. We append to the system a state $\rho$ prepared in the code $\cF$. The resulting state is stabilized by ${\sf QRM}(r,m+1)$. We can measure any additional stabilizers and used their associated pure errors to restore their +1 value in order to restore a given gauge. The logical operators of ${\sf QRM}(r,m)$ acting on the first block are preserved by this procedure provided that they do not conflict with the gauge choice.

\end{document}